# Application of Fractional Fourier Transform in Cepstrum Analysis

K. H. Miah*, R. H. Herrera, M. van der Baan, and M. D. Sacchi

University of Alberta, Edmonton, Canada

kmiah@ualberta.ca

**Summary**

Source wavelet estimation is the key in seismic signal processing for resolving subsurface structural properties. Homomorphic deconvolution using cepstrum analysis has been an effective method for wavelet estimation for decades. In general, the inverse of the Fourier transform of the logarithm of a signal's Fourier transform is the cepstral domain representation of that signal. The convolution operation of two signals in the time domain becomes an addition in the cepstral domain. The fractional Fourier transform (FRFT) is the generalization of the standard Fourier transform (FT). In an FRFT, the transformation kernel is a set of linear chirps whereas the kernel is composed of complex sinusoids for the FT. Depending on the fractional order, signals can be represented in multiple domains. This gives FRFT an extra degree of freedom in signal analysis over the standard FT. In this paper, we have taken advantage of the multidomain nature of the FRFT and applied it to cepstral analysis. We term this combination the Fractional-Cepstrum (FC). We derive the real FC formulation, and give an example using wavelets to show the multidomain representation of the traditional cepstrum with different fractional orders of the FRFT.

**Introduction**

Seismograms are often considered to be the convolution of the impulse response of the geology and the source excitation. Separating the source wavelet from the seismic trace is the key in predicting subsurface reflectivity. Since a convolution operation becomes an addition in the homomorphic domain, the wavelet becomes decoupled from the trace and maps into different regions in the cepstral domain (Ulrych et al., 1972; Otis and Smith, 1977). The presence of different types of noise and overlapping components make this separation a difficult problem to resolve.

Like the FT, the FRFT (Namias, 1980; Almeida, 1994; Ozaktas et al., 1996) has complex coefficients, and can be used to obtain both magnitude and phase information of a signal. The motivation of combining the FRFT with homomorphic transform is to take advantage of the multidomain property of the FRFT, and subsequently to have different cepstral representations of a signal. This new cepstral representation could potentially perform better in wavelet estimation than the standard FT based cepstrum.

The concept of the FRFT in the traditional cepstrum analysis for chirp signal processing was briefly discussed in the paper by Zhe, Hongyu, and Tianshuang (2004). In this paper, we first defined FRFT, and the convolution operation suitable for cepstrum implementation in the FRFT domain. Then a real Fractional-Cepstrum formula is derived by applying FRFT in the standard cepstrum. The fractional order parameter in the FRFT determines the different 'cepstrum' representations in the Fractional-Cepstrum domain. Depending on the fractional order, different real FC realizations of a synthetic trace are observed.

We think that the combination of FRFT and cepstrum will find applications in geophysical signal processing. Note, the notations 'Fractional-Cepstrum' and 'FC' are interchangeable throughout this paper.

**Theories of FRFT, Convolution and Cepstrum**

In the Fractional-Cepstrum, the forward and inverse Fourier transforms are replaced with their corresponding FRFTs. The main difference in defining the cepstrum using the FRFT comes from the fact that the conventional convolution operation defined for FT does not hold for the FRFT. Thus, the FRFT of the convolution of two time domain functions is not equal to the multiplication of their respective FRFT. This implies that to compute the complex cepstrum an additional term will be modulating the phase spectrum, while the real cepstrum approach remains the same.

*FRFT Principles:* The FRFT of a function $x(t)$ for the fractional order $\alpha \neq 0, \pi/2, \pi$ can be written as (Almeida, 1994):

$$X_\alpha(u) = \sqrt{\frac{1-j\cot\alpha}{2\pi}} \int_{-\infty}^{+\infty} \exp\left(j\left(\frac{\cot\alpha}{2}t^2 - ut\csc\alpha + \frac{\cot\alpha}{2}u^2\right)\right) x(t)\,dt. \tag{1}$$

Different formulations of product and convolution operations for the FRFT have been investigated (Almeida, 1997; Zayed, 1998; Singh and Saxena, 2010). Considering implementation flexibility, we have chosen Zayed's (1998) definition of the convolution of two functions for the FRFT in this paper:

$$h(t) = x(t) \otimes y(t), \tag{2}$$

$$h(t) = \sqrt{\frac{1-j\cot\alpha}{2\pi}} e^{-j\frac{\cot\alpha}{2}t^2} (\tilde{x} * \tilde{y}), \tag{3}$$

where $\otimes$ indicates convolution for the FRFT, $\tilde{x} = x(t)e^{j\frac{\cot\alpha}{2}t^2}$ and $\tilde{y} = y(t)e^{j\frac{\cot\alpha}{2}t^2}$. Let $H_\alpha$, $X_\alpha$ and $Y_\alpha$ denote the FRFT of $h$, $x$ and $y$, respectively. Then

$$H_\alpha(u) = X_\alpha(u) Y_\alpha(u) e^{-j\frac{\cot\alpha}{2}u^2}. \tag{4}$$

*Cepstrum Representation:* Let $s(t)$ represent a time-varying signal formed by the traditional convolution of two different functions, $w(t)$ and $r(t)$. Then real-valued cepstrum representations of $s(t)$ with the FT (Ulrych et al., 1972; Oppenheim, Kopec, and Tribolet, 1976) and FRFT can be derived as:

*Traditional FT based real Cepstrum*

$$s(t) = w(t) * r(t)$$
$$\Downarrow \text{FFT}$$
$$S(f) = W(f)R(f)$$
$$\Downarrow \text{LN}$$
$$\ln[S(f)] = \ln[W(f)] + \ln[R(f)]$$
$$\Downarrow \text{IFFT}$$
$$\hat{s}(t) = \hat{w}(t) + \hat{r}(t)$$

*FRFT based real Cepstrum*

$$s(t) = w(t) \otimes r(t)$$
$$\Downarrow \text{FRFT}_\alpha$$
$$S_\alpha(u) = W_\alpha(u) R_\alpha(u) e^{-j\frac{\cot\alpha}{2}u^2}$$
$$|S_\alpha(u)|e^{j\angle S_\alpha(u)} = |W_\alpha(u)|e^{j\angle W_\alpha(u)} \cdot |R_\alpha(u)|e^{j\angle R_\alpha(u)} \cdot e^{-j\frac{\cot\alpha}{2}u^2}$$
$$\Downarrow \text{LN } \textit{of magnitude terms}$$
$$\ln[|S_\alpha(u)|] = \ln[|W_\alpha(u)|] + \ln[|R_\alpha(u)|] + \ln[1]$$
$$\Downarrow \text{FRFT}_{-\alpha}$$
$$\hat{s}_\alpha(t) = \hat{w}_\alpha(t) + \hat{r}_\alpha(t)$$

## Example of Fractional-Cepstrum (FC)

A true wavelet is generated by superposition of two Ricker wavelets of different frequencies and phase rotations. Then the wavelet is convolved with a reflectivity series to form a synthetic trace (Fig. 1). Next, the real-valued Fractional-Cepstrum for the wavelet, reflectivity series, and the trace are plotted for various fractional orders $\alpha$ (Fig. 2) to depict the multidomain representation of the FRFT.

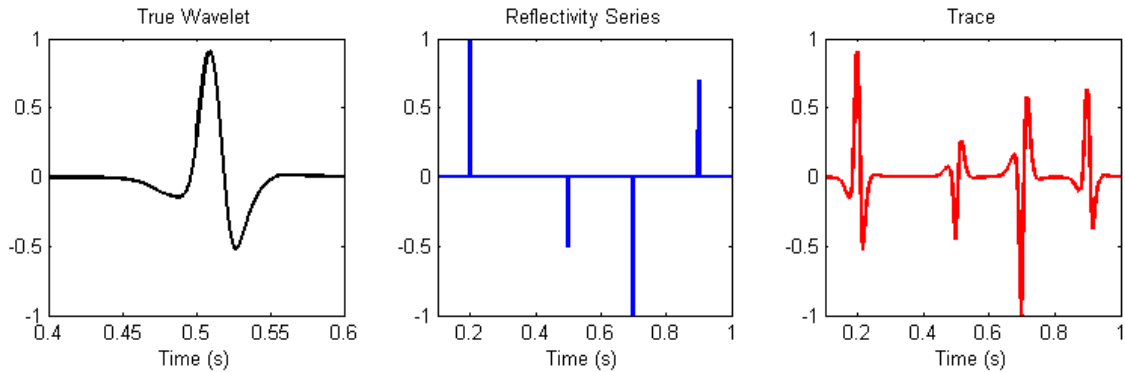

Figure 1: Synthetic trace is generated by convolving the wavelet with a reflectivity series for the FC analysis.

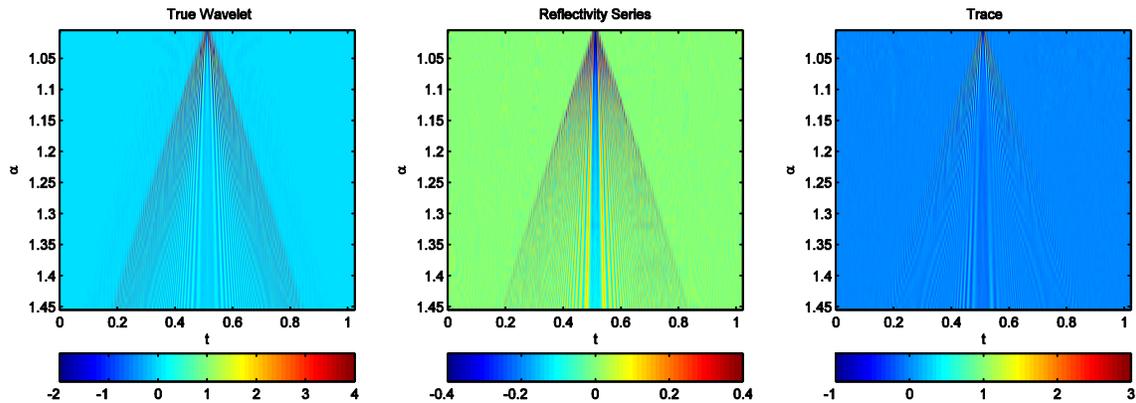

Figure 2: Real Fractional-Cepstrum realizations for different fractional order α.

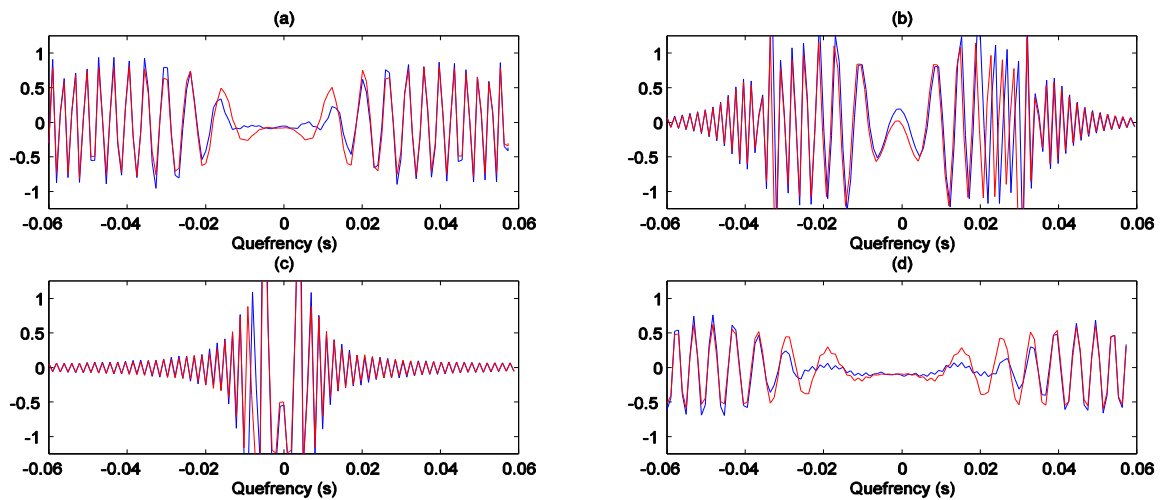

Figure 3: Real Fractional-Cepstrum realizations of the sum (wavelet + reflectivity) **red** and overall (trace) **blue** were compared, for 4 different fractional orders: (a) α=0.9, (b) α=0.95, (c) α=0.99, and (d) α=1.15.

In Figure 2, each image is the cepstral domain representation of the signal for varying fractional order. Each row of data represents the 'cepstrum' of the signal for a specific $\alpha$. It is to mention that $\alpha = 1$ corresponds to the FT based 'true' cepstrum. Figure 3, show that the summation of the real-valued Fractional-Cepstrum representation of any two signals is equivalent to the Fractional-Cepstrum of the two signals combined. It can be inferred from these comparisons (between $\hat{s}(t)$ and $\hat{w}(t) + \hat{r}(t)$) that even if we change the fractional order, the real-valued Fractional-Cepstrum obeys the fundamental principle of cepstrum – the convolution operation defined for the FRFT in the time-domain becomes an addition in the real-valued Fractional-Cepstrum domain.

## Conclusions

We have derived the real Fractional-Cepstrum formula by applying the FRFT instead of FT in the traditional Cepstrum analysis. This derivation of real-valued FC can form the basis for the complex FC analysis. We have not addressed phase unwrapping and related issues in this paper. It is the subject for future research in conjunction with the complex FC. Like the FRFT, we think that the caveat of the potential application of FC in seismic data processing application would be to find a set of optimum fractional orders that would perform better than that of the Fourier transform. Since the FRFT kernel is based on a set of linear chirp functions, signals characterized by strong chirp-like properties may be ideally suited for FC analysis. We hope this paper will serve as food for thought for researchers involved in geophysical signal processing using cepstral analysis.

## Acknowledgements

KHM thanks corporate sponsors of SAIG (Signal and Imaging Group) for financial support and RHH to the sponsors of BLISS (BLind Identification of Seismic Signals).